\documentclass[manuscript]{aastex}

\usepackage{graphicx}
\usepackage[varg]{txfonts}
\usepackage{lscape,longtable}
\usepackage{float}
\usepackage[latin1]{inputenc}
\usepackage[normalem]{ulem}

\usepackage{natbib}
\newcommand{\FeH}{$\mbox{[Fe/H]}$} 
\newcommand{\AB}[2]{$\mbox{[#1/#2]}$} 
\newcommand{\XFe}[1]{$\mbox{[#1/Fe]}$} 
\newcommand{\XH}[1]{$\mbox{[#1/H]}$} 

\newcommand{\FeHeq}[1]{$\mbox{[Fe/H]}={#1}$}       
\newcommand{\FeHlt}[1]{$\mbox{[Fe/H]}<{#1}$}       
\newcommand{\ABlt}[3]{$\mbox{[#1/#2]}<{#3}$}       
\newcommand{\ABle}[3]{$\mbox{[#1/#2]}\le{#3}$}     
\newcommand{\tefft}{$T_{\mbox{\scriptsize eff}}$}  
\newcommand{\teffm}{T_{\mbox{\scriptsize eff}}}    

\newcommand{\pun}[1]{\mbox{\rm\,#1}} 
\newcommand{\logg}{\ensuremath{\log g}}

\newcommand{\mlp}{\ensuremath{\alpha_{\mathrm{MLT}}}}
\newcommand{\moh}{\ensuremath{[\mathrm{M/H}]}}

\newcommand{\Teff}{\ensuremath{T_{\mathrm{eff}}}}
\newcommand{\miiid}{\ensuremath{\left\langle\mathrm{3D}\right\rangle}}

\slugcomment{Fluorine Abundances of Galactic Low-Metallicity Giants}

\shortauthors{Li et al.}

\begin{document}

\title{Fluorine Abundances of Galactic Low-Metallicity Giants}

\author{H. N. Li}
\affil{Key Lab of Optical Astronomy, National Astronomical Observatories,
		Chinese Academy of Sciences,
              	A20 Datun Road, Chaoyang, Beijing 100012, China}
\email{lhn@nao.cas.cn}

\author{H.-G. Ludwig}
\affil{Zentrum f{\"u}r Astronomie der Universit{\"a}t Heidelberg, Landessternwarte,
		K{\"o}nigstuhl 12, D-69117 Heidelberg, Germany}
\email{hludwig@lsw.uni-heidelberg.de}

\author{E. Caffau}
\affil{Zentrum f{\"u}r Astronomie der Universit{\"a}t Heidelberg, Landessternwarte,
		K{\"o}nigstuhl 12, D-69117 Heidelberg, Germany}
\email{ecaffau@lsw.uni-heidelberg.de}


\author{N. Christlieb}
\affil{Zentrum f{\"u}r Astronomie der Universit{\"a}t Heidelberg, Landessternwarte,
		K{\"o}nigstuhl 12, D-69117 Heidelberg, Germany}
\email{N.Christlieb@lsw.uni-heidelberg.de}

\and

\author{G. Zhao}
\affil{Key Lab of Optical Astronomy, National Astronomical Observatories,
		Chinese Academy of Sciences,
              	A20 Datun Road, Chaoyang, Beijing 100012, China}
\email{gzhao@nao.cas.cn}

\received{2012 October 4}
\accepted{2013 January 15}

\begin{abstract}
With abundances and 2-$\sigma$ upper limits of fluorine (F) in seven metal-poor field giants,
nucleosynthesis of stellar F at low metallicity is discussed.
The measurements are derived from the HF(1-0) R9 line at 23358\,{\AA}
using near-infrared $K$-band high-resolution spectra obtained with CRIRES at the Very Large
Telescope.
The sample reaches lower metallicities than previous studies on F of field giants,
ranging from \FeHeq{-1.56} down to $-$2.13. Effects of three-dimensional model atmospheres
on the derived F and O abundances are quantitatively estimated
and shown to be insignificant for the program stars.
The observed F yield in the form of \AB{F}{O} is compared with two sets of
Galactic chemical evolution models, which quantitatively
demonstrate the contribution of Type
II supernova (SN II) $\nu$-process and asymptotic giant branch/Wolf¨CRayet stars.
It is found that at this low-metallicity region,
models cannot well predict the observed distribution of \AB{F}{O},
while the observations are better fit by models
considering an SN II $\nu$-process
with a neutrino energy of $E_{\nu}=3\times10^{53}$ erg.
Our sample contains HD~110281, a retrograde orbiting low-$\alpha$ halo star,
showing a similar F evolution as globular clusters. This supports the theory
that such halo stars are possibly accreted from dwarf galaxy progenitors
of globular clusters in the halo.
\end{abstract}


\section{Introduction}

Although fluorine is a very interesting element, its chemical evolution is not yet well understood.
F has only one stable, yet fragile isotope, $^{19}$F,
which is easily destroyed in the stellar interiors by the most abundant elements,
H and He, by means of the reactions $^{19}$F(p,$\alpha$)$^{16}$O
and $^{19}$F($\alpha$,p)$^{22}$Ne. Theoretically, in order to explain the production of F,
three sites have been proposed to explain the mechanism that enables it
to escape from the hot stellar interior after it has been created, including
Type II supernovae (SN II), asymptotic giant branch (AGB) stars, and Wolf-Rayet (WR) stars.
It is predicted that SN II produce $^{19}$F primarily as a result of neutrino spallation
\citep{Woosley1988Nature,Woosley1995ApJS,Heger2005Ph,Kobayashi2011ApJL}.
For AGB stars, \citet{Forestini1992AA} proposed that $^{19}$F is synthesized and
then dredged up to the surface during the He-burning thermal pulses.
In WR stars, $^{19}$F is probably produced during the He-burning phase and then
ejected into space by strong stellar winds before it is destroyed \citep{Meynet2000AA}.

There are very few atomic or molecular lines of F that are suitable for stellar abundance analysis.
Current studies on F have focused on the measurement of the single HF molecular line
at 23358\,{\AA} in cool stars. \citet{Jorissen1992AA} put up a pioneering work in the field
by measuring the first F abundance of a number of giants with near-solar metallicities.
Through the past decade, more studies on F nucleosynthesis have been carried out,
aiming at various types of objects, including solar- and moderate-metallicity global cluster giants
\citep{Cunha2003AJ,Werner2005AA,Smith2005ApJ}, giants in the Galactic bulge \citep{Cunha2008ApJ,Uttenthaler2008ApJ},
AGB stars \citep{Abia2009ApJ,Abia2010ApJ}, and main-sequence dwarfs in the solar neighborhood \citep{Recio-Blanco2012AA}.
Besides, detailed exploration of the chemical evolution of F has been carried out
targeting chemically peculiar objects, such as cool extreme He stars \citep{Pandey2006ApJ},
C-enhanced stars \citep{Schuler2007ApJL,Lucatello2011ApJ},
Ba stars \citep{Alves-Brito2011AA}, and even an extragalactic C-star \citep{Abia2011ApJ}.

With new nucleosynthesis yields and updated observations, chemical evolution models
can now predict more details of the nucleosynthesis of F, but only the contribution of AGB stars
to the F production has been observationally confirmed \citep{Werner2005AA,Abia2010ApJ,Abia2011ApJ}.
The chemical evolution model for MilkyWay-like disk galaxies of \citet{Renda2004MNRAS}
suggests a scenario where AGB stars only began to contribute
to the production of F after $\sim$ 1~Gyr. This model states that AGB stars are at present
the major contributors for the production of F; while in the early Galaxy,
only SN~II contribute to the F abundance, and provide a solution to the discrepancy between
models and observed F abundance in Galactic giant in the intermediate-metallicity range.
A more recent chemical evolution model targeting fluorine by \citet{Kobayashi2011ApJL} adopts the $\nu$-process
yields and yields of AGB stars with $1-7$\,$M_{\odot}$, based on the infall and
star formation history in \citet{Kobayashi2011MNRAS}.
The model predicts a \AB{F}{O} plateau at low metallicity depending on the neutrino luminosity.
The included $\nu$-process enhances the predicted F abundances at higher metallicities,
and helps to fit the high F abundances observed in field and bulge giants.
Both models are in good agreement with observations in regions with solar or
moderately low metallicities,
indicating the importance for constraining the theoretical yields
at low metallicity for further scrutiny and future observations.
\citet{Alves-Brito2012AA} filled the gap at the low-metallicity area with M22 data,
but the complex chemical enrichment history of this globular cluster leads to a large dispersion
(0.6\,dex) of the F abundances,
thus does not follow the trend of other globular clusters or being explained by a general model.
Hence, the metal-poor field giants in our observed sample, which have metallicities of \FeHlt{-1.5},
become very interesting in allowing us to map the pristine F abundance in earlier phases
of the Galactic halo evolution.

Observations and data reduction, including the determination of stellar parameters
and abundance analysis, are addressed in Sections~\ref{sec:obs} through ~\ref{sec:analysis}.
The three-dimensional (3D) effects on the abundances are described in Section~\ref{sec:3D}.
In Section~\ref{sec:discussion}, our results are discussed and interpreted,
and a brief summary is given in Section~\ref{sec:conclusion}.

\section{CRIRES Observations}\label{sec:obs}

\begin{figure}
\epsscale{.90}
\plotone{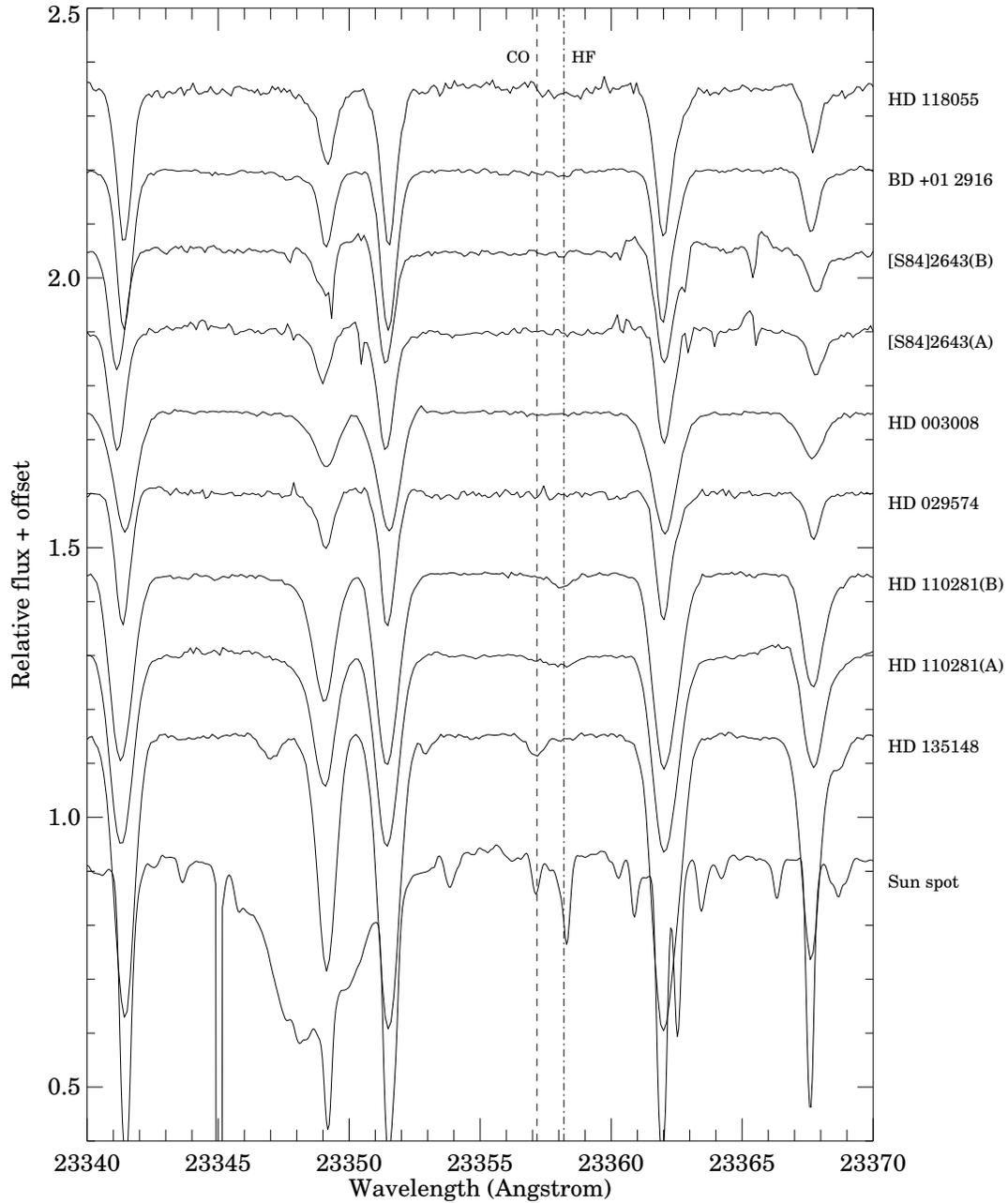} 
\caption{Overview of the HF-region spectra for all exposures of the programs stars.
The position of the HF R9 line and the closeby C$^{17}$O are indicated by dash-dotted and dashed lines respectively.
A $K$-band Sun spot spectrum in this area is also plotted for comparison.}\label{fig:overview}
\end{figure}

Seven metal-poor field giants were observed with CRIRES \footnote{http:www.eso.org/instruments/crires}
(Cryogenic high-resolution Infrared Echelle Spectrograph; \citealt{Kaeufl2004SPIE})
at Very Large Telescope UT1 in a number of service runs from 2008 May through 2009 March.
The slit width was set to 0".2, yielding the maximum resolution of 100,000
($\sim$ 3 \pun{km s$^{-1}$}). The four CCD chips cover the wavelength ranges
from 22993 to 23118\,{\AA}, 23152 to 23273\,{\AA}, 23304 to 23420\,{\AA}, and 23450 to 23560\,{\AA}.
Chip 3 includes the HF(1-0) R9 line which, in cool stars,
has been verified to be most reliable for F abundance determination.
The integration time was 30\,s or 45\,s for each nodding position per setting,
and for two of the program stars, two exposures were obtained.
A hot standard star with similar air mass was observed immediately either beforehand or afterward.
The raw frames of science and standard objects were reduced with the CRIRES pipeline (v2.0.0),
and after that, the science spectra were divided by corresponding standard star spectra
to correct for the telluric lines and the illumination pattern, using the IRAF task Telluric.
The typical signal-to-noise (S/N) ratio of the sample spectra around 23358 {\AA} is
higher than 200.
Spectra of the HF line area for all exposures of our program stars are shown in Figure~\ref{fig:overview}.

\section{Atmospheric Parameters}\label{sec:atmos_param}

The effective temperature \tefft\  and the surface gravity \logg\
can be determined either spectroscopically or photometrically.
The microturbulent velocity $\xi$ is normally derived by
removing any trend of abundances with equivalent widths of Fe lines.
However, the fact that there is no Fe feature in the sample spectra
within the coverage of CRIRES observation makes it difficult to independently
determine the atmospheric parameters with the IR spectra alone.
Thus, the stellar parameters for the sample stars were adopted from previously
published data.
Out of a number of measurements, we relied on the work of \citet{Shetrone1996AJ},
which includes all seven objects in our sample and thus provides
a uniform set of atmospheric parameters. Moreover, it contains measurements of
equivalent widths.

The model atmospheres adopted by \citet{Shetrone1996AJ} are taken from the
grid of \citet{Bell1976AAS}, which is now outdated.  Therefore, we have
adopted the atmospheric parameters of \citet{Shetrone1996AJ} as the initial
values, and further adjusted them with local thermal equilibrium (LTE) one-dimensional (1D)
models computed with ATLAS9 \citep{Kurucz1994CD}.  The adjustment followed the
traditional spectroscopic method. \tefft\ was adjusted to make
\ion{Fe}{1} lines with low excitation potential yield the same abundances as those
with high excitation potential; the microturbulent velocity was adjusted until
strong \ion{Fe}{1} lines yield same abundances as weak \ion{Fe}{1} lines;
and \logg\ was adjusted so that abundances from \ion{Fe}{1} and \ion{Fe}{2} lines
were forced to agree with each other.
The equivalent width (EW) of the Fe lines were directly
adopted from the measurements of \citet{Shetrone1996AJ}.  The adjusted
parameters do not differ much from \citet{Shetrone1996AJ}'s values,
with a difference being 10$\pm$45 K for \tefft, $-$0.03$\pm$0.08 for \logg\,
and $-$0.02$\pm$0.03 for \FeH.

Similarly, with the EW measurement and the linelist of \citet{Shetrone1996AJ},
abundances of O, Mg, Al, and Eu (which cannot be determined from
CRIRES spectra since no corresponding line features are present in the $K$-band spectra)
were also re-calculated, using the ATLAS9 atmospheric models.
The re-derived abundances show overall agreement with Shetrone's values,
e.g., with a difference of 0.10$\pm$0.08 for O, 0.10$\pm$0.06 for Mg, $-$0.09$\pm$0.07 for Al,
and 0.06$\pm$0.06 for Eu, respectively. 
The adopted atmospheric parameters and abundances, based on the optical data,
are listed in Table~\ref{tab:param_abun}.

\section{Abundance Analysis}\label{sec:analysis}

\begin{figure}
\epsscale{.90}
\plotone{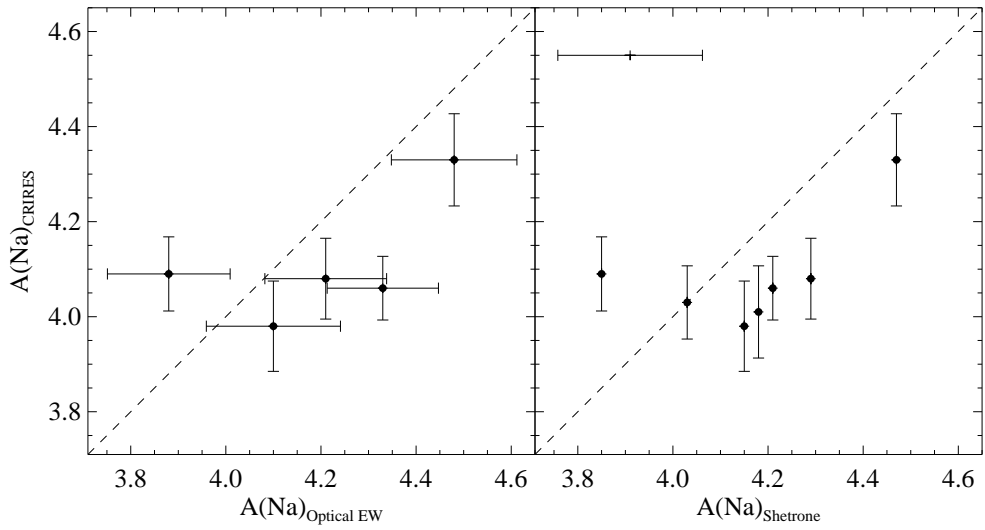} 
\caption{Na abundances derived from CRIRES spectra are compared with those re-derived
using \citet{Shetrone1996AJ}'s EWs measured from optical spectra (left),
and the original results provided by \citet{Shetrone1996AJ} (right).
Two objects with no EWs available from the literature are not included in the left plot.}
\label{fig:Na_abundance}
\end{figure}

Using the spectrum synthesis code MOOG \citep{Sneden1973ApJ} and
model atmospheres calculated with ATLAS9 \citep{Kurucz1994CD},
LTE Na, C, and F abundances (and upper limits) were obtained
by comparison with CRIRES spectra.
The mixing-length parameter was set to $\mlp=1.25$,
and no convective overshooting was adopted.
The wavelength range of the CRIRES spectra covers a few Phillips system C$_{2}$ lines;
however, since our sample stars are of rather low-metallicities
and without carbon enhancement (except for HD~135148), no C$_{2}$ line was detected.
Therefore, we relied on the O abundances obtained as previously described,
and derived the C abundances by fitting several $^{12}$C$^{16}$O first overtone
vibrational-rotational lines (2--0 and 3--1 vibration series).
Although \citet{Shetrone1996AJ} also provides measurements of Na abundances,
which was determined as an average of the abundances
derived from EWs for the 6154\,{\AA} and 6160 \,{\AA} NaI lines and
from synthetic spectrum fits to the 5682 \,{\AA} and 5688 \,{\AA} NaI lines. 
there are two objects without Na EW measurements in the literature.
However, there are two Na lines located at 23379\,{\AA},
which are strong and clean enough to measure the Na abundance with the IR spectra.
The left plot of Figure~\ref{fig:Na_abundance} compares Na abundances
determined from CRIRES spectra and re-derived from the EWs measured by \citet{Shetrone1996AJ}
for the five stars with EW measurements, and the right plot compares the CRIRES Na abundances
with the average Na abundances given by \citet{Shetrone1996AJ}.
The measurements agree within their uncertainties,
and the CRIRES values are adopted for the following discussion.

\begin{figure*}[htbp]
\epsscale{.90}
\plotone{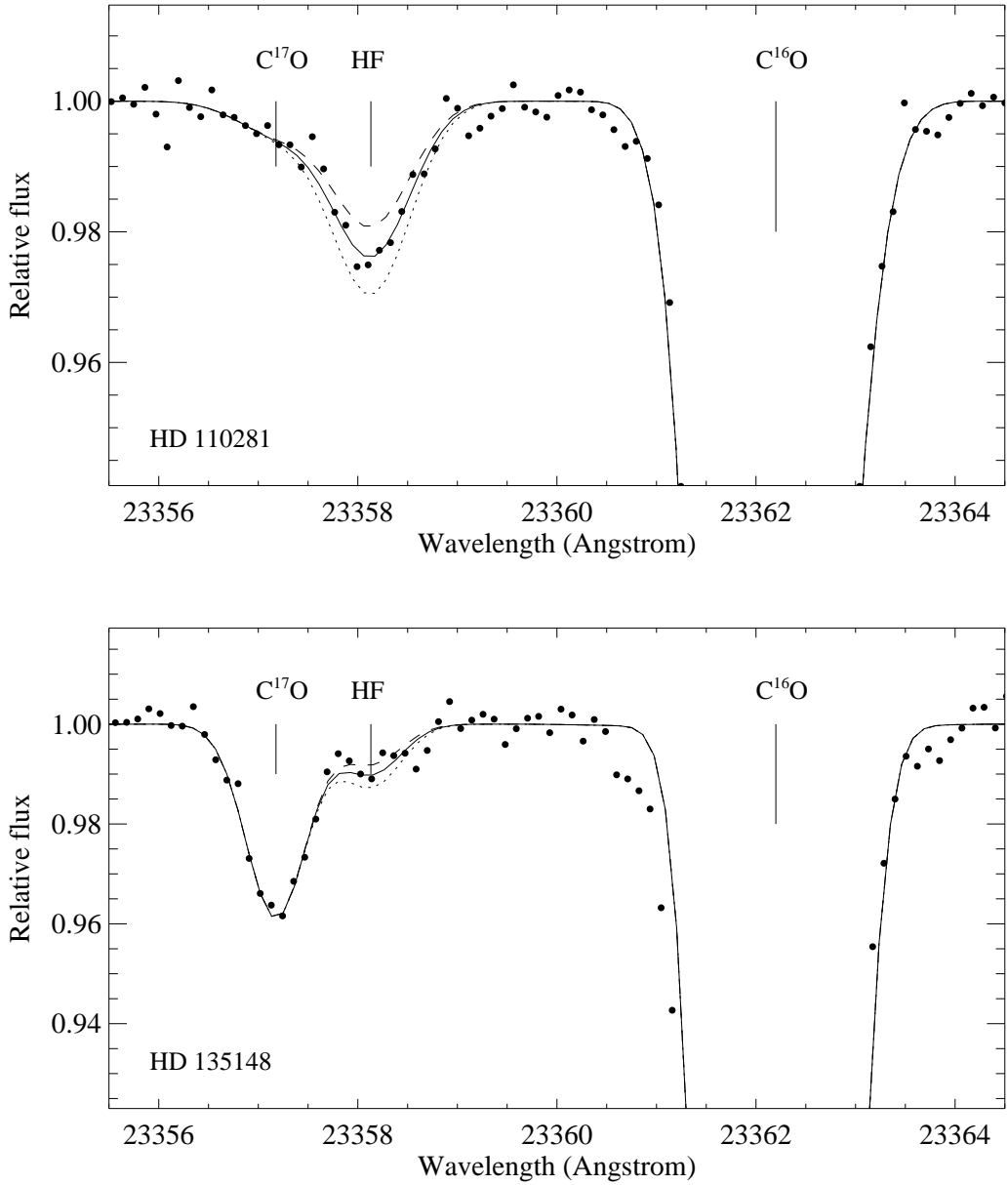} 
\caption{Sample spectra with spectral synthesis of HF line at 23358\,{\AA}
of HD~110281 and HD~135148, the two program stars with clear HF detections.
Black dots are observed data points. Solid lines refer to the synthetic spectra
with the adopted A(F), with dashed lines and dotted lines referring to synthetic spectra with $\Delta$A(F)=$\pm$ 0.1~dex.
Note that for the CH~star HD~135148, the C$^{17}$O line to the blue side of the HF line
is clearly detected and also fitted.}\label{fig:spectra}
\end{figure*}

The F abundances of our sample stars were determined by fitting the unblended HF (1--0) R9 line at $23358$\,{\AA}.
Out of the seven sample objects, clear HF R9 features were detected
in two of them, HD~110281 and HD~135148, as shown in Figure~\ref{fig:spectra}.
Note that HD~135148 is a spectroscopic binary with a companion star
with the approximate mass of a white dwarf (e.g., \citealt{Carney2003AJ}),
it is thus not likely that the detected HF feature is reflecting the closeby CO lines
from the companion stellar spectra. To confirm this, we have also checked the
auto-correlation of the spectra, and found no significant second peak
associated with a companion star.
For the remaining five spectra, 1-$\sigma$ upper limits of F abundances were obtained,
adopting the statistical error of EW based on the classical formula of \citet{Cayrel1988IAUS}:
\begin{equation}
\langle \Delta W ^{2} \rangle ^{1/2} \simeq 1.6 (w \Delta x)^{1/2} \epsilon
\label{equa:error}
\end{equation}
where $w$ is the FWHM of the line; $\Delta x$ refers to the sampling step of the spectra,
and $\epsilon$ to the reciprocal spectral S/N in the case of normalized spectra.
Considering the fact that the HF line is weak in the CRIRES spectra,
we can estimate their expected FWHM from weak CO lines close to the HF line.
The S/N of the spectra were calculated from several continuum regions
on the same CCD as the HF feature. Note that except for the spectroscopic binary HD~135148,
two other objects, HD~003008 and HD~110281 clearly show broader lines than other program stars.
Spectroscopic and photometric observations by \citet{Carney2003AJ} have concluded
that the line-broadening of these two stars is caused by fast rotation rather than stellar companions,
and also proposed the idea that these two rapid rotators show enhanced mass loss
which is aided by the rotation, or by the addition of both orbital energy and angular momentum
from an accreted planet to the stellar envelope.
We have also investigated the detectability of HF lines with various EWs/upper limits,
e.g., the difference in A(F) between the two exposures of HD~110281
corresponds to a 1-2 m{\AA} difference in EW, which is on the same level as
the 1-$\sigma$ upper limit.
We have adopted 2-$\sigma$ upper limits of A(F) in the following discussion,
to ensure the possible detection considering the typical S/N and line widths of our program stars.

For the only C-enhanced star HD~135148, the ratio of $^{16}$O/$^{17}$O was derived
based on three $^{12}$C$^{17}$O lines in our spectra, at 23347\,{\AA},
23357\,{\AA}, and 23379\,{\AA}. A $^{16}$O/$^{17}$O ratio of 140$\pm$50 was derived.
This oxygen isotope ratio is higher than normal at such a metallicity,
if one only considers the nucleosynthesis of massive stars.
However, since $^{17}$O can be overproduced by AGB stars, and AGB stars can be
the typical companion star that transfers nucleosynthesis products to the C-enhanced star,
the high ratio of $^{16}$O/$^{17}$O can be explained by this scenario.
Note that the $^{12}$C$^{17}$O line at 23357\,{\AA}
is located right to the blue wing of the HF R9 line, but is not at all affecting
the derived F abundance, as testified in early work like \citet{Abia2009ApJ}.
The fitted $^{12}$C$^{17}$O line is also shown in Figure~\ref{fig:spectra}.
For the synthetic fitting, we have adopted the Na linelist from \citet{Kurucz1994CD},
the CO linelist from \citet{Goorvitch1994ApJS}, and the HF linelist from \citet{Cunha2003AJ}.
Validity of the linelist has been checked by fitting the $K$-band spectra
of $\alpha$ Boo from the atlas of \citet{Hinkle1995PASP}.

The uncertainties of the abundance of F and other elements mainly comes from two aspects:
the statistical error in the line equivalent widths, and the uncertainties
of the atmospheric parameters.
To estimate the former, statistical errors of the EW measurements were calculated adopting Equation~\ref{equa:error},
and the latter were determined by individually varying stellar parameters
to measure their separate effects on the derived abundances.
Typically, e.g., for HD~110281, the 1-$\sigma$ EW error leads to an uncertainty of
0.02\,dex of the F abundance,
$\Delta \teffm=+100$~K leads to an increase of +0.25 of the F abundance,
$\Delta \log g = +0.5$ produces $\Delta A(F)=-0.02$,
and $\Delta \xi = +0.5$~\pun{km s$^{-1}$} gives a negligible $\delta A(F) < 0.001$.
As for the Na abundance, the typical sensitivities to EW errors and atmospheric parameters
are $\Delta A(Na)=+0.06$, $+$0.07, $-$0.03, and $-$0.01 respectively.
The resulting uncertainties are then summed in quadrature to obtain the total statistical errors.
For stars with more than one exposure, i.e., HD~110281 and $[$S84$]$~2643,
abundances for the exposure with higher S/N was adopted,
and the differences between the derived element abundances from different
exposures turned out to be negligible.
The derived abundances of C, O, F, Na, Mg, Al, and Eu are listed in Table~\ref{tab:param_abun}.

\section{3D effects}\label{sec:3D}

\subsection{Abundance Corrections for the HF~R9 and Forbidden O Lines}

It is well-known that atmospheres of metal-depleted, late-type stars are prone
to exhibit significant deviations from radiative equilibrium
\citep{Asplund1999AA}. In particular, abundances derived from molecular
species can experience large corrections
\citep[e.g.,][]{Hernandez2008AA}. From this perspective it is desirable
to quantify the 3D effects -- at least for fluorine and oxygen. Unfortunately,
we have no 3D models available at the rather low gravities of our program
stars. To nevertheless derive an estimate of the abundance corrections, we
took four 3D models from the CIFIST atmosphere grid \citep{Ludwig2009MmSAI} at
higher gravity and inspected the trend of the abundance corrections with
changing \Teff\ and \logg. When calculating the abundance corrections, we
followed the methodology described in \cite{Caffau2011SoPh}. In the 3D models we
assumed a solar chemical composition scaled to a metallicity of $\moh=-2.0$, with
an enhancement of the $\alpha$-elements by $+0.4\pun{dex}$.

Table~\ref{tab:3dcorr} summarizes the results for the HF~R9 line, and the two
forbidden oxygen lines at 6300\pun{\AA} and 6363\pun{\AA}. The listed 3D
abundance corrections are evaluated for weak lines (negligible saturation) --
a good approximation considering the typical line strength. When running the 3D models,
sufficient statistics were gathered so that residual statistical uncertainties
on the abundance corrections are smaller than the accuracy to which they are
given in the table. For completeness, we mention that in the 1D hydrostatic
comparison models we assumed a mixing-length parameter of $\mlp=1.0$ (close to
what was adopted in the ATLAS9 models), and in the related 1D spectral
synthesis calculations a microturbulence of $\xi=2.0\pun{km s$^{-1}$}$. However, both
aspects have hardly any impact on the result. The corrections in
Tab.~\ref{tab:3dcorr} suggest that for the low-gravity giants of our sample,
minor 3D effects are to be expected. Only at gravities $\logg > 2$ we expect
substantial corrections ($>0.1\pun{dex}$) on the F abundance at $\moh=-2$. For
the HF line, mainly horizontal inhomogeneities (as opposed to differences in
the vertical stratification) are the cause of the changes of its strength in
3D compared to 1D models.

Digressing for the moment, we note that the abundance corrections obtained for
the forbidden oxygen lines are not fully compatible with the corrections
obtained by \cite{Collet2007AA}. These authors obtained $\approx
-0.1\pun{dex}$ at $\Teff=5050\pun{K}$, $\logg=2.2$, $\moh=-2.0$, and
$\mbox{[O/Fe]}=+0.5$. Even considering that the atmospheric parameters are not
exactly the same in their and our models, the corrections of Collet and
collaborators appear noticeably larger. This is also the case for
other species \citep[see][]{Ivanauskas2010nuco, Dobrovolskas2010nuco} and
the reason is not yet understood.

Nevertheless, we proceed under the assumption that 3D corrections are
insignificant for the F and O abundances we derived for our program stars.

\subsection{Convective Line Shifts of HF and CO Lines}

\begin{figure}
\epsscale{0.8}
\plotone{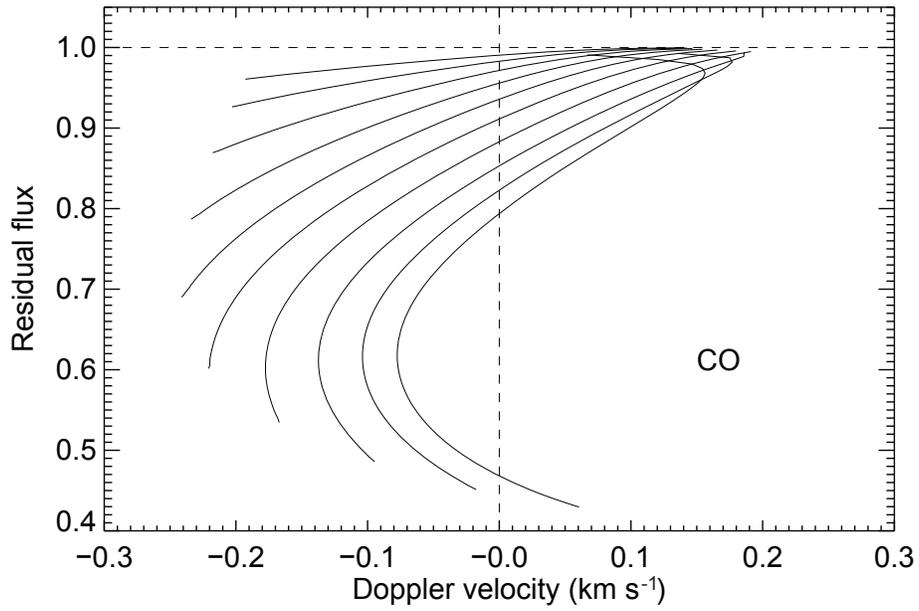} 
\caption{Line bisectors of typical CO lines (solid lines) of different strength in the HF
  region. The line synthesis was done at constant C/O ratio. Zero Doppler
  velocity corresponds to the laboratory wavelength.
}\label{fig:bisCO}
\end{figure}

\begin{figure}
\epsscale{0.8}
\plotone{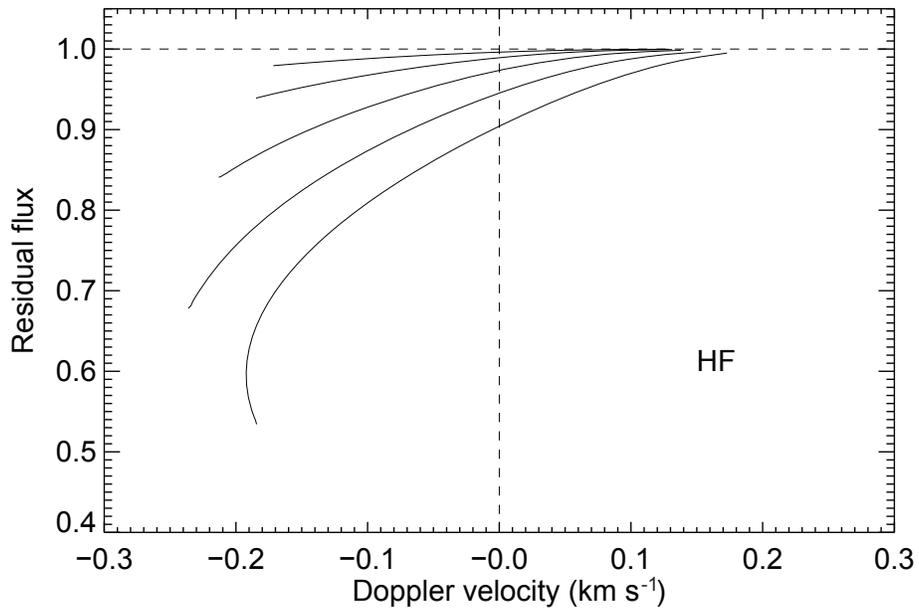} 
\caption{Line bisectors of the HF~R9 line (solid lines) of different
  strength. Zero Doppler velocity corresponds to the laboratory wavelength
}\label{fig:bisHF}
\end{figure}

We used the cores of strong CO lines as synthesized by a 1D spectral synthesis
code to correct for the imperfections of the wavelength calibration provided
by the CRIRES reduction pipeline. One might ask whether convective line shifts
alter the wavelength position of the cores significantly with respect to their
laboratory wavelength. Again, we do not have fully adequate 3D models
available for our target stars. We nevertheless used the 3D model from the
CIFIST grid which gets closest ($\Teff=3886\pun{K}$, $\logg=1.0$,
$\moh=-2.0$) to calculate line bisectors of the HF~R9 line and a prototypical
CO line in the HF region. We varied the strength of the lines by
(artificially) varying the oscillator strength; in the case of the CO line
this means that the variation was done at fixed C/O
ratio. Figures~\ref{fig:bisCO} and~\ref{fig:bisHF} illustrate the outcome. The
perhaps surprising result is that the magnitude of the convective shifts is
very modest, not exceeding 250\pun{m s$^{-1}$}. The cores of strong CO lines in our
target stars have typically a residual flux around 0.7, i.e., the HF line is
weak. Differential shifts between their cores are even significantly less,
unlikely to exceed 100\pun{m s$^{-1}$} corresponding to $\approx 8\pun{m\AA}$ at the
wavelength of the CRIRES spectra -- much smaller than the line width. We
conclude that our wavelength correction procedure should produce reliable
line positions for the HF and CO lines, as can be seen in Figure~\ref{fig:overview}.

\section{Discussion}\label{sec:discussion}

\begin{figure}
\epsscale{.8}
\plotone{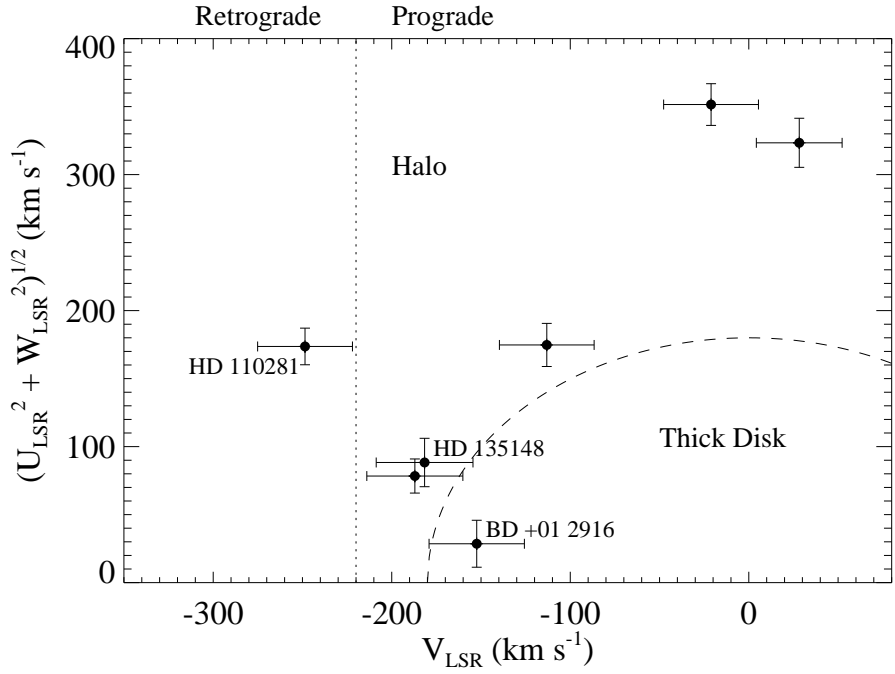} 
\caption{Toomre diagram of the program stars. The long-dashed line corresponds
to $V_{total}$ = 180 km s$^{-1}$. The short-dashed line indicates zero rotation in the Milky Way.
Two objects with HF detections, HD~110281 (with retrograde rotation) and
HD~135148 (the CH-star), and BD~+01~2916 (identified as a thick disk star)
are marked in the plot.}\label{fig:Toomre}
\end{figure}

\begin{figure}
\epsscale{1.1}
\plotone{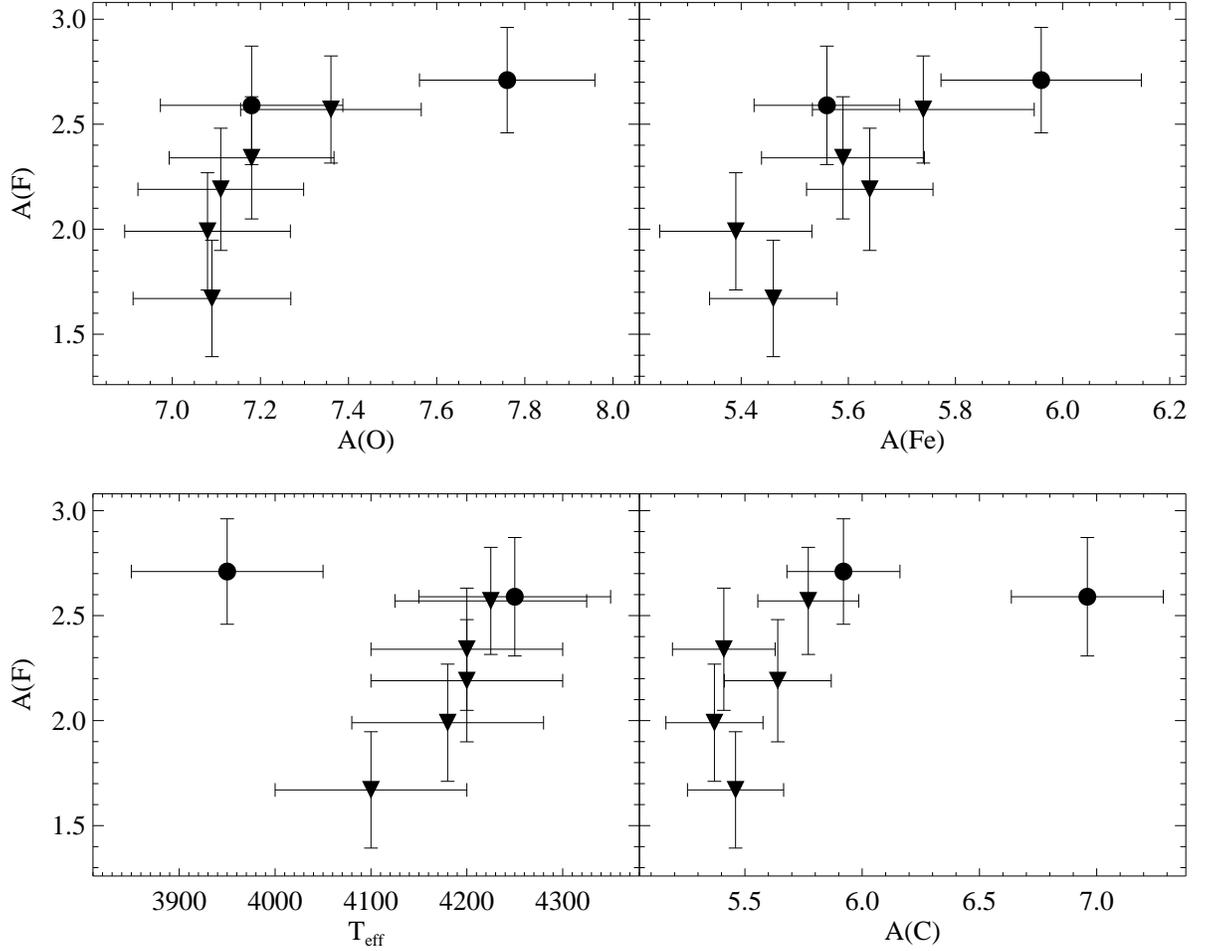} 
\caption{Variances of the F abundance along \tefft, abundances of C, O and Fe.
The filled circles with error bars refer to HD~110281 and HD~135148,
and all upside-down filled triangles correspond to the program stars
with 2-$\sigma$ upper limits of F for the cases in which the HF line is not detected.
1-$\sigma$ errors are overplotted to the observed data.}
\label{fig:F_Teff_C_O_Fe}
\end{figure}

With the measurements of F abundances and upper limits of the program stars,
it is now possible to probe the behavior of F evolution of field giants with low metallicities.
Figure~\ref{fig:Toomre} displays the kinematic properties of our program stars.
Please note that the radial velocity and proper motion data are directly obtained from previous literatures as listed in the database of SIMBAD \footnote{http://simbad.u-strasbg.fr/simbad/}.
As can be seen, one of the stars with clear HF detections, HD~110281,
is identified as a halo star with retrograde rotation,
and another program star, BD~+01~2916, is classified as a thick disk star.
The sensitivities of the F abundances are checked along temperatures,
and abundances of C, O, and Fe, as shown in Figure~\ref{fig:F_Teff_C_O_Fe}.
Relevant correlations or dependencies have been addressed in a number of articles,
including observations \citep{Jorissen1992AA,Cunha2003AJ,Abia2009ApJ},
as well as theoretical predictions from Galactic chemical evolution models
such as \citet{Renda2004MNRAS} and \citet{Kobayashi2011ApJL}.
We would expect for giant stars to find a correlation between the abundance of F
and the abundances of O and Fe, and that the F abundances could be enhanced by
stars exhibiting C-enhancement.
It is quite likely that potential trends as expected are present in the comparisons,
but we cannot draw firm conclusions on this since our sample size is small and we have
obtained mostly upper limits only.
Nevertheless, it is clear that HD~110281, the object with the highest F abundance,
is the star with the lowest \Teff\ in our sample,
and showing the highest metallicity (Fe and O).
The HF molecule formation is very sensitive to temperature,
hence \Teff\  is expected to set the level of upper limits on the F abundance.
The other star with a HF-detection, HD~135148, is showing notably
higher C abundances compared to other program stars,
since it is a CH-star as shown by \citet{Carney2003AJ} and \citet{Shetrone1999PASP}.
The observed trend of the F abundances with metallicity tracers
suggest that O and Fe may be used as a reference on the detection limit of F;
however, it is not clear to us whether there is a lower limit of F detection
determined by the stellar metallicity, because \citet{Alves-Brito2012AA}
has studied several M22 giants with metallicities (and stellar parameters)
that are comparable to our sample but most of their giants' spectra show measurable HF features.
Note that these spectra based on which the \citet{Alves-Brito2012AA} derive the F abundances
do not present higher resolution or S/N than our sample spectra;
therefore, it is suspected that the different detection fraction of HF features
between the two samples are possibly due to the fact that one is in clusters
while the other is in the field.

\begin{figure}
\epsscale{1.1}
\plotone{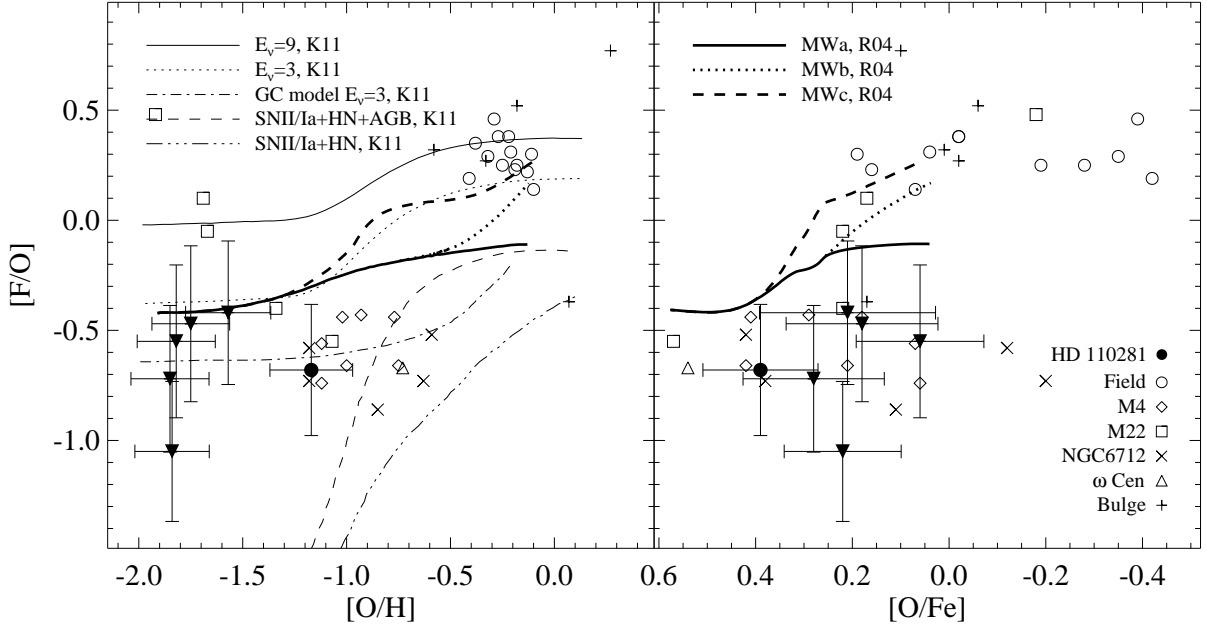} 
\caption{The F yields in the form of \AB{F}{O} as a function of \AB{O}{H} and \AB{O}{Fe}.
Models of Renda et al. (2004; thick lines) and Kobayashi et al. (2011a; thin lines) are shown,
with line identifications described in the plot and text. Observations of various giants are also included:
open circles for field K and M giants from \citet{Cunha2003AJ}, and \citet{Cunha2005ApJ};
diamonds, squares, crosses, and triangles for globular cluster giants from M4 \citep{Smith2005ApJ},
M22 \citep{Alves-Brito2012AA}, NGC6712 \citep{Yong2008ApJ}, and $\omega$ Cen \citep{Cunha2003AJ}, respectively; pluses for bulge giants from \citet{Cunha2008ApJ}. 
Filled symbols are the same as in Figure~\ref{fig:F_Teff_C_O_Fe}, except that the CH-star HD~135148
is not included in the plot due to its peculiarly enhanced F abundances as explained in the context.
All abundances of models and observations are calibrated to the same set of solar abundances
from \citet{Anders1989GeCoA}.}\label{fig:FO_OH}
\end{figure}

In Figure~\ref{fig:FO_OH}, \AB{F}{O} abundance ratios treated as the indicator of the F yield of the sample,
are plotted against \AB{O}{H} and \AB{O}{Fe} and compared with model predictions and previous observations.
Two sets of models are selected, including Renda et al. (2004, R04 hereafter) which for the first time
quantitatively demonstrated the contribution of AGB and WR stars to the F production,
and the latest model by Kobayashi et al. (2011a, hereafter K11), who calculate
the contribution of the $\nu$-process of SN II and hypernovae (HN) to the nucleosynthesis of fluorine.
Three model variants are constructed in R04: MWa (thick solid line) with SN II as the only source of $^{19}$F,
MWb (thick dotted line) including yields from both SN II and WR stars,
and MWc (thick dashed line) including all three sources of SN II, WR, and AGB stars.
For all three cases, R04 have considered that neutrino spallation is largely dominant
in fluorine production, and they adopted the SN II yields of \citet{Woosley1995ApJS},
which assumes a total energy in neutrinos of $E_{\nu}=3\times10^{53}$ erg.
As for K11, five cases are included:
evolution with neither AGBs nor $\nu$-process (SN II/Ia + HN, triple-dot-dashed line),
chemical evolution with AGB contributions (SN II/Ia + HN + AGB, dashed line),
the model for globular clusters with $\nu$-process (dash-dotted line),
$\nu$-process of SN II and HN (dash-dotted and solid lines) with neutrino luminosity
of $E_{\nu}=3\times10^{53}$ erg and $E_{\nu}=9\times10^{53}$ erg respectively.
Note, that K11 models do not include yields of WR stars, because the contribution
by WR to F may be reduced by including rotation in the stellar models.
For more details of the models, we refer the interested readers to R04 and K11.

As can be seen in Figure~\ref{fig:FO_OH}, for all cases taking $\nu$-process into account,
all models predict a plateau of \AB{F}{O} at low metallicity
with \ABle{O}{H}{-1.2}, as proposed by K11. The $\nu$-process remarkably enhances the level of F production
compared to models without $\nu$-process and indicates the dominant contribution from neutrino spallation
to producing F in the early stage of Galactic evolution.
For all four model variants with the same neutrino energies of $E_{\nu}=3\times10^{53}$ erg,
the predictions of the F production at the plateau of the models of R04 and K11 agree
well with each other.
For a better judgment of the significance of the differences between our observations and the model predictions,
1-$\sigma$ error bars are also included to the upper limits of \AB{F}{O} of the sample.
As C-enhanced stars are believed to be s-process enhanced and producers of F,
the CH-star HD~135148 shows the highest F abundances as shown in Table~\ref{tab:param_abun},
thus is not included in Figure~\ref{fig:FO_OH} which intends to compare with the general trend.
Also noted, due to its different kinematical properties from the rest of the sample,
HD~110281 (shown in filled circle) may not follow the Galactic halo evolution,
hence should be separately considered when compared with models.
Except for these two objects, all upper limits of the other five objects in our sample
(as shown with lower triangular)
are located around the predicted plateau of R04 and the SN II $\nu$-process model of K11,
while well below the HN model of K11 with $E_{\nu}=9\times10^{53}$ erg. 
The general trend of the observed \AB{F}{O} with increasing metallicity,
and the apparent discrepancy between the observed \AB{F}{O} and predictions by
models without neutrino spallation, suggest that a contribution of the $\nu$-process is needed
to explain the early production of F in the Galactic halo.
The program stars with intermediate F content in the sample tend to follow the
evolution model for SN II $\nu$-process of K11 and the early phase of R04 models,
which predict a similar ''plateau'', but program stars with low F upper limits
are distributed at relatively lower level of \AB{F}{O}, and cannot be explained
by these models.

Our observed F abundances are also compared with previous observations,
including giants of globular clusters \citep{Cunha2003AJ,Smith2005ApJ,Yong2008ApJ,Alves-Brito2012AA},
the Milky Way bulge \citep{Cunha2008ApJ}, as well as K and M field giants \citep{Cunha2003AJ,Cunha2005ApJ}.
As shown in the left panel of Figure~\ref{fig:FO_OH},
for globular cluster giants with comparable metallicity, e.g., M22 \citep{Alves-Brito2012AA},
the detected F level shows clear difference from our program stars,
which could be explained by the different chemical enrichment history of field and cluster stars,
and/or the wide range of chemical abundance variations of M22 due to its complex formation history.
Previously available data of field giants covered only metallicities close to solar,
thus they show a notably higher F abundances than our metal-deficient field giants.
Apart from the field and globular cluster halo giants, Milky Way bulge giants \citep{Cunha2008ApJ}
are also compared. They are showing remarkably higher F abundances with a large scatter,
which is difficult to explain by yields of current models.
This also implies a different evolution of this central component of the Milky Way.
However, it is interestingly noticed that HD~110281 follows very similar trend of \AB{F}{O}
as giants in globular clusters such as M4, NGC6712, and $\omega$ Cen.
Note in Table~\ref{tab:param_abun} that this object also shows rather low Na and Mg abundances,
as well as being on a retrograde orbit in Figure~\ref{fig:Toomre},
thus it is very likely a "low-$\alpha$" halo star as defined by \citet{Nissen2010AA,Nissen2011AA}.
As they have found based on the abundance and kinematic analysis,
the less-bound low-$\alpha$ stars are probably accreted from dwarf galaxies
with relatively slow star-formation,
and some of them may be associated with the progenitor galaxy of globular clusters,
including $\omega$ Cen, which could refer to a similar or relevant chemical evolution,
hence explain the similarity in F evolution of HD~110281 and the compared globular cluster giants.
Future observations of abundances of F and other elements of similar objects would be very interesting
and important to obtain a deeper understanding of the origin of the dual halo of the Galaxy.

The timescale of SN II is much shorter than that of AGB or WR stars, which naturally explains
that SN II dominate the area with lower \AB{F}{O} and higher \XFe{O},
whereas AGB stars start to notably contribute after \XH{O} $> -1.5$ and become more important
at higher metallicity and higher \AB{F}{O}. Therefore, the F abundance is a good clock
to distinguish the contribution from low-mass AGB stars and SN II.
It is true that based on our comparisons, none of the included models could well fit or predict
the observations at lower metallicities, neither for field giants nor globular clusters.
But it is also true that both the R04 and K11 models are originally aimed at better fitting
the observational data of globular clusters at intermediate metallicities
and giants toward solar metallicity due to the availability of observational data.
This is also the reason why the observed F production from AGB stars is better explained,
while there is still little known about the early evolution of F dominated by massive stars
or the different F evolution of halo field giants and globular clusters stars.
Future observations of field stars at low metallicities
are crucial for further constraining  models for such early stages of the
Galactic chemical evolution,
e.g., the neutrino luminosity released from SN II, and the initial conditions
of the chemical evolution models for the older halo.

\section{Summary and conclusion}\label{sec:conclusion}

With CRIRES observations of seven halo field giants at \FeHlt{-1.5} and \ABlt{O}{H}{-1.1},
and abundances of O, Mg, Al, and Eu determined in optical spectra from the literature,
we have investigated the nucleosynthesis of their F content,
and for the first time, explored the evolution of F in field giants in the low-metallicity regime.
Among the program stars, the HF R9 feature is detected in two cases, HD~110281 and HD~135148,
and 2-$\sigma$ upper limits are estimated for the F abundance of the other stars.
Using CIFIST model atmospheres, the 3D effect on the F and O abundance
with comparable stellar parameters to the program stars are estimated,
showing that for our sample, the corrections are negligible.
The sensitivity of the F abundance/upper limits are also checked against \Teff, metallicities, O,
and C. A correlation can be found between the upper limits of F and \Teff,
which confirms the expectation that more stringent limits can be obtained for cooler stars,
and C-enhancement or higher metallicity are also enhancing the F abundance or upper limits.
Galactic chemical evolution models by \citet{Renda2004MNRAS} and \citet{Kobayashi2011ApJL}
predict a similar F production at low metallicity, and are compared with observations.
It is suggested that the $\nu$-process is necessary to explain the F production
at the early stages of Galactic evolution, but the predicted \AB{F}{O} ratios
do not fit well the observed distribution, whereas the observed trend is closest to
the SN II model of \citet{Kobayashi2011ApJL} and early phases of \citet{Renda2004MNRAS} models
with a low total energy in neutrinos, $E_{\nu}=3\times10^{53}$ erg.
Combined with kinematics, HD~110281 is identified
as a low-$\alpha$ star \citep{Nissen2010AA} with retrograde rotation.
The consistent F nucleosynthesis of this star with the globular cluster halo stars
support the scenario that this kind of objects may be accreted from nearby dwarf galaxies
and enriched by early metal-deficient SN II. Future observations and modifications on models
on relevant elements of the Galactic halo are important to achieve a better understanding
of the above-mentioned results and remaining puzzles.

\acknowledgments
This work was supported by NSFC grant Nos. 11233004, 11103030, and 10903012,
the Global Networks program of Universit{\"a}t
Heidelberg, and grant CH~214/5-1 of Deutsche Forschungsgemeinschaft (DFG).
We further acknowledge support by Sonderforschungsbereich SFB~881 ''The Milky
Way System'' (subprojects A4 and A5) of the DFG.

\bibliographystyle{apj} 
\bibliography{fluorine} 

\begin{thebibliography}{45}
\expandafter\ifx\csname natexlab\endcsname\relax\def\natexlab#1{#1}\fi

\bibitem[{{Abia} {et~al.}(2011){Abia}, {Cunha}, {Cristallo}, {de Laverny},
  {Dom{\'{\i}}nguez}, {Recio-Blanco}, {Smith}, \& {Straniero}}]{Abia2011ApJ}
{Abia}, C., {Cunha}, K., {Cristallo}, S., {de Laverny}, P., {Dom{\'{\i}}nguez},
  I., {Recio-Blanco}, A., {Smith}, V.~V., \& {Straniero}, O. 2011, \apjl, 737,
  L8

\bibitem[{{Abia} {et~al.}(2009){Abia}, {Recio-Blanco}, {de Laverny},
  {Cristallo}, {Dom{\'{\i}}nguez}, \& {Straniero}}]{Abia2009ApJ}
{Abia}, C., {Recio-Blanco}, A., {de Laverny}, P., {Cristallo}, S.,
  {Dom{\'{\i}}nguez}, I., \& {Straniero}, O. 2009, \apj, 694, 971

\bibitem[{{Abia} {et~al.}(2010){Abia}, {Cunha}, {Cristallo}, {de Laverny},
  {Dom{\'{\i}}nguez}, {Eriksson}, {Gialanella}, {Hinkle}, {Imbriani},
  {Recio-Blanco}, {Smith}, {Straniero}, \& {Wahlin}}]{Abia2010ApJ}
{Abia}, C., {et~al.} 2010, \apjl, 715, L94

\bibitem[{{Alves-Brito} {et~al.}(2011){Alves-Brito}, {Karakas}, {Yong},
  {Mel{\'e}ndez}, \& {V{\'a}squez}}]{Alves-Brito2011AA}
{Alves-Brito}, A., {Karakas}, A.~I., {Yong}, D., {Mel{\'e}ndez}, J., \&
  {V{\'a}squez}, S. 2011, \aap, 536, A40

\bibitem[{{Alves-Brito} {et~al.}(2012){Alves-Brito}, {Yong}, {Mel{\'e}ndez},
  {V{\'a}squez}, \& {Karakas}}]{Alves-Brito2012AA}
{Alves-Brito}, A., {Yong}, D., {Mel{\'e}ndez}, J., {V{\'a}squez}, S., \&
  {Karakas}, A.~I. 2012, \aap, 540, A3

\bibitem[{{Anders} \& {Grevesse}(1989)}]{Anders1989GeCoA}
{Anders}, E., \& {Grevesse}, N. 1989, \gca, 53, 197

\bibitem[{{Asplund} {et~al.}(1999){Asplund}, {Nordlund}, {Trampedach}, \&
  {Stein}}]{Asplund1999AA}
{Asplund}, M., {Nordlund}, {\AA}., {Trampedach}, R., \& {Stein}, R.~F. 1999,
  \aap, 346, L17

\bibitem[{{Bell} {et~al.}(1976){Bell}, {Eriksson}, {Gustafsson}, \&
  {Nordlund}}]{Bell1976AAS}
{Bell}, R.~A., {Eriksson}, K., {Gustafsson}, B., \& {Nordlund}, A. 1976, \aaps,
  23, 37

\bibitem[{{Caffau} {et~al.}(2011){Caffau}, {Ludwig}, {Steffen}, {Freytag}, \&
  {Bonifacio}}]{Caffau2011SoPh}
{Caffau}, E., {Ludwig}, H.-G., {Steffen}, M., {Freytag}, B., \& {Bonifacio}, P.
  2011, \solphys, 268, 255

\bibitem[{{Carney} {et~al.}(2003){Carney}, {Latham}, {Stefanik}, {Laird}, \&
  {Morse}}]{Carney2003AJ}
{Carney}, B.~W., {Latham}, D.~W., {Stefanik}, R.~P., {Laird}, J.~B., \&
  {Morse}, J.~A. 2003, \aj, 125, 293

\bibitem[{{Cayrel}(1988)}]{Cayrel1988IAUS}
{Cayrel}, R. 1988, in IAU Symposium, Vol. 132, The Impact of Very High S/N
  Spectroscopy on Stellar Physics, ed. {G.~Cayrel de Strobel \& M.~Spite}, 345

\bibitem[{{Collet} {et~al.}(2007){Collet}, {Asplund}, \&
  {Trampedach}}]{Collet2007AA}
{Collet}, R., {Asplund}, M., \& {Trampedach}, R. 2007, \aap, 469, 687

\bibitem[{{Cunha} \& {Smith}(2005)}]{Cunha2005ApJ}
{Cunha}, K., \& {Smith}, V.~V. 2005, \apj, 626, 425

\bibitem[{{Cunha} {et~al.}(2008){Cunha}, {Smith}, \& {Gibson}}]{Cunha2008ApJ}
{Cunha}, K., {Smith}, V.~V., \& {Gibson}, B.~K. 2008, \apjl, 679, L17

\bibitem[{{Cunha} {et~al.}(2003){Cunha}, {Smith}, {Lambert}, \&
  {Hinkle}}]{Cunha2003AJ}
{Cunha}, K., {Smith}, V.~V., {Lambert}, D.~L., \& {Hinkle}, K.~H. 2003, \aj,
  126, 1305

\bibitem[{{Dobrovolskas} {et~al.}(2010){Dobrovolskas}, {Kucinskas}, {Ludwig},
  {Caffau}, {Klevas}, \& {Prakapavicius}}]{Dobrovolskas2010nuco}
{Dobrovolskas}, V., {Kucinskas}, A., {Ludwig}, H.~G., {Caffau}, E., {Klevas},
  J., \& {Prakapavicius}, D. 2010, in Nuclei in the Cosmos.

\bibitem[{{Forestini} {et~al.}(1992){Forestini}, {Goriely}, {Jorissen}, \&
  {Arnould}}]{Forestini1992AA}
{Forestini}, M., {Goriely}, S., {Jorissen}, A., \& {Arnould}, M. 1992, \aap,
  261, 157

\bibitem[{{Gonz{\'a}lez Hern{\'a}ndez} {et~al.}(2008){Gonz{\'a}lez
  Hern{\'a}ndez}, {Bonifacio}, {Ludwig}, {Caffau}, {Spite}, {Spite}, {Cayrel},
  {Molaro}, {Hill}, {Fran{\c c}ois}, {Plez}, {Beers}, {Sivarani}, {Andersen},
  {Barbuy}, {Depagne}, {Nordstr{\"o}m}, \& {Primas}}]{Hernandez2008AA}
{Gonz{\'a}lez Hern{\'a}ndez}, J.~I., {et~al.} 2008, \aap, 480, 233

\bibitem[{{Goorvitch}(1994)}]{Goorvitch1994ApJS}
{Goorvitch}, D. 1994, \apjs, 95, 535

\bibitem[{{Heger} {et~al.}(2005){Heger}, {Kolbe}, {Haxton}, {Langanke},
  {Mart{\'{\i}}nez-Pinedo}, \& {Woosley}}]{Heger2005Ph}
{Heger}, A., {Kolbe}, E., {Haxton}, W.~C., {Langanke}, K.,
  {Mart{\'{\i}}nez-Pinedo}, G., \& {Woosley}, S.~E. 2005, Physics Letters B,
  606, 258

\bibitem[{{Hinkle} {et~al.}(1995){Hinkle}, {Wallace}, \&
  {Livingston}}]{Hinkle1995PASP}
{Hinkle}, K., {Wallace}, L., \& {Livingston}, W. 1995, \pasp, 107, 1042

\bibitem[{{Ivanauskas} {et~al.}(2010){Ivanauskas}, {Kucinskas}, {Ludwig}, \&
  {Caffau}}]{Ivanauskas2010nuco}
{Ivanauskas}, A., {Kucinskas}, A., {Ludwig}, H.~G., \& {Caffau}, E. 2010, in
  Nuclei in the Cosmos.

\bibitem[{{Jorissen} {et~al.}(1992){Jorissen}, {Smith}, \&
  {Lambert}}]{Jorissen1992AA}
{Jorissen}, A., {Smith}, V.~V., \& {Lambert}, D.~L. 1992, \aap, 261, 164

\bibitem[{{Kaeufl} {et~al.}(2004){Kaeufl}, {Ballester}, {Biereichel},
  {Delabre}, {Donaldson}, {Dorn}, {Fedrigo}, {Finger}, {Fischer}, {Franza},
  {Gojak}, {Huster}, {Jung}, {Lizon}, {Mehrgan}, {Meyer}, {Moorwood}, {Pirard},
  {Paufique}, {Pozna}, {Siebenmorgen}, {Silber}, {Stegmeier}, \&
  {Wegerer}}]{Kaeufl2004SPIE}
{Kaeufl}, H.-U., {et~al.} 2004, in Society of Photo-Optical Instrumentation
  Engineers (SPIE) Conference Series, Vol. 5492, Society of Photo-Optical
  Instrumentation Engineers (SPIE) Conference Series, ed. {A.~F.~M.~Moorwood \&
  M.~Iye}, 1218--1227

\bibitem[{{Kobayashi} {et~al.}(2011{\natexlab{a}}){Kobayashi}, {Izutani},
  {Karakas}, {Yoshida}, {Yong}, \& {Umeda}}]{Kobayashi2011ApJL}
{Kobayashi}, C., {Izutani}, N., {Karakas}, A.~I., {Yoshida}, T., {Yong}, D., \&
  {Umeda}, H. 2011{\natexlab{a}}, \apjl, 739, L57

\bibitem[{{Kobayashi} {et~al.}(2011{\natexlab{b}}){Kobayashi}, {Karakas}, \&
  {Umeda}}]{Kobayashi2011MNRAS}
{Kobayashi}, C., {Karakas}, A.~I., \& {Umeda}, H. 2011{\natexlab{b}}, \mnras,
  414, 3231

\bibitem[{{Kurucz}(1994)}]{Kurucz1994CD}
{Kurucz}, R. 1994, Solar abundance model atmospheres for 0,1,2,4,8 km/s.~Kurucz
  CD-ROM No.~19.~ Cambridge, Mass.: Smithsonian Astrophysical Observatory,
  1994., 19

\bibitem[{{Lucatello} {et~al.}(2011){Lucatello}, {Masseron}, {Johnson},
  {Pignatari}, \& {Herwig}}]{Lucatello2011ApJ}
{Lucatello}, S., {Masseron}, T., {Johnson}, J.~A., {Pignatari}, M., \&
  {Herwig}, F. 2011, \apj, 729, 40

\bibitem[{{Ludwig} {et~al.}(2009){Ludwig}, {Caffau}, {Steffen}, {Freytag},
  {Bonifacio}, \& {Ku{\v c}inskas}}]{Ludwig2009MmSAI}
{Ludwig}, H.-G., {Caffau}, E., {Steffen}, M., {Freytag}, B., {Bonifacio}, P.,
  \& {Ku{\v c}inskas}, A. 2009, \memsai, 80, 711

\bibitem[{{Meynet} \& {Arnould}(2000)}]{Meynet2000AA}
{Meynet}, G., \& {Arnould}, M. 2000, \aap, 355, 176

\bibitem[{{Nissen} \& {Schuster}(2010)}]{Nissen2010AA}
{Nissen}, P.~E., \& {Schuster}, W.~J. 2010, \aap, 511, L10

\bibitem[{{Nissen} \& {Schuster}(2011)}]{Nissen2011AA}
---. 2011, \aap, 530, A15

\bibitem[{{Pandey}(2006)}]{Pandey2006ApJ}
{Pandey}, G. 2006, \apjl, 648, L143

\bibitem[{{Recio-Blanco} {et~al.}(2012){Recio-Blanco}, {de Laverny}, {Worley},
  {Santos}, {Melo}, \& {Israelian}}]{Recio-Blanco2012AA}
{Recio-Blanco}, A., {de Laverny}, P., {Worley}, C., {Santos}, N.~C., {Melo},
  C., \& {Israelian}, G. 2012, \aap, 538, A117

\bibitem[{{Renda} {et~al.}(2004){Renda}, {Fenner}, {Gibson}, {Karakas},
  {Lattanzio}, {Campbell}, {Chieffi}, {Cunha}, \& {Smith}}]{Renda2004MNRAS}
{Renda}, A., {et~al.} 2004, \mnras, 354, 575

\bibitem[{{Schuler} {et~al.}(2007){Schuler}, {Cunha}, {Smith}, {Sivarani},
  {Beers}, \& {Lee}}]{Schuler2007ApJL}
{Schuler}, S.~C., {Cunha}, K., {Smith}, V.~V., {Sivarani}, T., {Beers}, T.~C.,
  \& {Lee}, Y.~S. 2007, \apjl, 667, L81

\bibitem[{{Shetrone}(1996)}]{Shetrone1996AJ}
{Shetrone}, M.~D. 1996, \aj, 112, 1517

\bibitem[{{Shetrone} {et~al.}(1999){Shetrone}, {Smith}, {Briley}, {Sandquist},
  \& {Kraft}}]{Shetrone1999PASP}
{Shetrone}, M.~D., {Smith}, G.~H., {Briley}, M.~M., {Sandquist}, E., \&
  {Kraft}, R.~P. 1999, \pasp, 111, 1115

\bibitem[{{Smith} {et~al.}(2005){Smith}, {Cunha}, {Ivans}, {Lattanzio},
  {Campbell}, \& {Hinkle}}]{Smith2005ApJ}
{Smith}, V.~V., {Cunha}, K., {Ivans}, I.~I., {Lattanzio}, J.~C., {Campbell},
  S., \& {Hinkle}, K.~H. 2005, \apj, 633, 392

\bibitem[{{Sneden}(1973)}]{Sneden1973ApJ}
{Sneden}, C. 1973, \apj, 184, 839

\bibitem[{{Uttenthaler} {et~al.}(2008){Uttenthaler}, {Aringer}, {Lebzelter},
  {K{\"a}ufl}, {Siebenmorgen}, \& {Smette}}]{Uttenthaler2008ApJ}
{Uttenthaler}, S., {Aringer}, B., {Lebzelter}, T., {K{\"a}ufl}, H.~U.,
  {Siebenmorgen}, R., \& {Smette}, A. 2008, \apj, 682, 509

\bibitem[{{Werner} {et~al.}(2005){Werner}, {Rauch}, \& {Kruk}}]{Werner2005AA}
{Werner}, K., {Rauch}, T., \& {Kruk}, J.~W. 2005, \aap, 433, 641

\bibitem[{{Woosley} \& {Haxton}(1988)}]{Woosley1988Nature}
{Woosley}, S.~E., \& {Haxton}, W.~C. 1988, \nat, 334, 45

\bibitem[{{Woosley} \& {Weaver}(1995)}]{Woosley1995ApJS}
{Woosley}, S.~E., \& {Weaver}, T.~A. 1995, \apjs, 101, 181

\bibitem[{{Yong} {et~al.}(2008){Yong}, {Mel{\'e}ndez}, {Cunha}, {Karakas},
  {Norris}, \& {Smith}}]{Yong2008ApJ}
{Yong}, D., {Mel{\'e}ndez}, J., {Cunha}, K., {Karakas}, A.~I., {Norris}, J.~E.,
  \& {Smith}, V.~V. 2008, \apj, 689, 1020

\end{thebibliography}

\clearpage

\clearpage

\clearpage

\clearpage

\clearpage

\clearpage

\clearpage

\clearpage



\clearpage

\begin{landscape}
\begin{table*}[htbp]
\begin{center}
\caption{Stellar Parameters and Abundances of the Sample.}\label{tab:param_abun}
\begin{tabular}{lccccccrcrccrr}
\\
\tableline\tableline
Star          &\tefft&\logg&\FeH&$\xi \pun{km s$^{-1}$}$&FWHM\tablenotemark{a}& S/N&A(C)&A(O)&A(F)   &A(Na)&A(Mg)&A(Al)&A(Eu)\\
\tableline
BD~$+$01~2916 &  4200& 0.10& $-$1.88& 2.10&0.709 & 357.3         &5.64&7.11&$<$2.19&4.08 &6.21 &4.46 &$-$1.13\\
HD~003008     &  4100& 0.10& $-$2.06& 1.90&1.030 & 591.8         &5.46&7.09&$<$1.67&3.98 &6.33 &4.21 &$-$1.21\\
HD~029574     &  4200& 0.10& $-$1.93& 1.75&0.658 & 226.8         &5.41&7.18&$<$2.34&4.01 &6.30 &4.40 &$-$0.95\\
HD~110281     &  3950& 0.20& $-$1.56& 2.43&0.996 & 462.5         &5.92&7.76&   2.71&4.33 &6.42 &4.70 &$-$0.46\\
HD~118055     &  4225& 0.70& $-$1.78& 1.63&0.687 & 163.6         &5.77&7.36&$<$2.57&4.03 &6.20 &4.58 &$-$0.86\\
HD~135148     &  4250& 0.70& $-$1.96& 2.40&0.899 & 360.2         &6.96&7.18&   2.59&4.06 &6.02 &4.59 &---\tablenotemark{b}\\
$[$S84$]$~2643&  4180& 0.40& $-$2.13& 1.65&0.735 & 316.3         &5.37&7.08&$<$1.99&4.09 &6.07 &4.50 &---\tablenotemark{b}\\
\tableline
\end{tabular}
\tablenotetext{a}{FWHM of the spectral lines of the sample stars (in m{\AA}), based on CO line measurements.}
\tablenotetext{b}{No Eu line or EW measurement is available for this object from \citet{Shetrone1996AJ}.}
\end{center}
\end{table*}
\end{landscape}

\clearpage

\begin{table}
\begin{center}
\caption{3D Abundance Corrections\tablenotemark{a} on the HF R9 and Forbidden Oxygen Lines.}\label{tab:3dcorr}
\begin{tabular}{ccrrrrrrc}
\\
\tableline\tableline
\Teff & \logg &
\multicolumn{2}{c}{HF R9} &
\multicolumn{2}{c}{[{O}\,{I}]\,6300\AA} &
\multicolumn{2}{c}{[{O}\,{I}]\,6363\AA} &
3D model\tablenotemark{b}\\
\tableline
3886 & 1.0 & -0.02 & -0.03 &  0.00 &  0.01 &  0.00 &  0.01 & d3t36g10mm20n02\\
4001 & 1.5 & -0.03 & -0.03 &  0.00 &  0.01 &  0.00 &  0.01 & d3t40g15mm20n01\\
4478 & 2.5 & -0.16 & -0.12 & -0.01 & -0.01 & -0.01 & -0.01 & d3t45g25mm20n02\\
5024 & 2.5 & -0.42 & -0.52 &  0.00 & -0.02 &  0.00 & -0.02 & d3t50g25mm20n01\\
\tableline
\end{tabular}
\tablenotetext{a}{\mbox{Left} columns give $\mathrm{3D}-\mathrm{1D}$, right columns
  $\mathrm{3D}-\miiid$, in dex.}
\tablenotetext{b}{All models assume $\moh=-2.0$, and $[\alpha/\mathrm{Fe}]=+0.4$.}
\end{center}
\end{table}

\end{document}